\documentclass{pasj01}
\Received{$\langle$August 4, 2016$\rangle$}
\Accepted{$\langle$September 8, 2016$\rangle$}
\Published{$\langle$publication date$\rangle$}
\SetRunningHead{Astronomical Society of Japan}{Usage of \texttt{pasj00.cls}}
\usepackage[dvips]{color}

\def\sgm{$\sigma$}
\def\pbeam{beam$^{-1}$}

\def\kms{km~s$^{-1}$}

\def\Vsys{$v_\mathrm{sys}$}

\def\wat{H$_2$O}
\def\Lsun{$L_{\odot}$}

\def\Msun{\mbox{$M_\odot$}}

\def\Reff{\mbox{$R_\mathrm{eff}$}}

\begin{document}

\title{A Massive Dense Gas Cloud close to the Nucleus of the Seyfert galaxy NGC\,1068}

\author{
Ray S. Furuya\altaffilmark{1} and
Yoshiaki~Taniguchi\altaffilmark{2}
}
\email{rsf@tokushima-u.ac.jp}
\altaffiltext{1}{Tokushima University, 1-1 Minami Jousanjima-machi, Tokushima 770-8502, Japan}
\altaffiltext{2}{The Open University of Japan, 2-11 Wakaba, Mihama-ku, Chiba 261-8586, Japan}

\maketitle

\begin{abstract}

Using the ALMA archival data of both $^{12}$CO\,(6--5) line and 689\,GHz continuum
emission towards the archetypical Seyfert galaxy, NGC\,1068, 
we identified a distinct continuum peak separated by 14\,pc
from the nuclear radio component S1 in projection.
The continuum flux gives a gas mass of $\sim 2 \times 10^5$ \Msun\
and bolometric luminosity of $\sim 10^8$ \Lsun, leading to a
star formation rate of  $\sim$ 0.1 \Msun\ yr$^{-1}$. 
Subsequent analysis on the line data suggest that the gas has a size of $\sim 10$\,pc,
yielding to mean H$_2$ number density of $\sim 10^5$ cm$^{-3}$.
We therefore refer to the gas as ``massive dense gas cloud":
the gas density is high enough to form a ``proto starcluster" whose
stellar mass of $\sim 10^4$ \Msun.
We found that the gas stands a unique position between 
galactic and extraglactic clouds
in the diagrams of
start formation rate (SFR) vs. gas mass proposed by Lada et al. and
surface density of gas vs. SFR density by Krumholz and McKee.
All the gaseous and star-formation properties  may be understood
in terms of the turbulence-regulated star formation scenario.
Since there are two stellar populations with the ages of 300\,Myr and 30\,Myr
in the 100 pc-scale circumnulear region, we discuss that 
NGC\,1068 has experienced at least three episodic star formation events
with a tendency that the inner star-forming region is the  younger. 
Together with several lines of evidence that the dynamics of the nuclear region
is decoupled from that of the entire galactic disk, we discuss that the gas inflow towards 
the nuclear region of NGC\,1068 may be driven by a past minor merger.
\end{abstract}

\section{Introduction} \label{s:intro}

NGC\,1068 is one of the nearest archetypical Seyfert galaxies in the nearby Universe
\citep{Sey43, KW74}, making it an ideal laboratory towards understanding
active galactic nuclei (AGNs) \citep{AM85}; the distance of 15.9 Mpc  \citep{KH13}
is adopted throughout this paper.
Therefore, a number of observational studies at various wavelengths
have been made to understand the nature of AGN phenomena in NGC\,1068
[e.g., \citet{Cecil02, SB12, Mezcua15, LR16, Wang12}].\par

Another important issue is the so-called starburst-AGN connection
since a number of Seyfert galaxies have  an intense 
circumnuclear ($\sim 100$\,pc scale)
star forming region around its AGN
[e.g., \citet{Sim80, Wilson91, SB96} and references therein]. 
Although there is no general consensus for this issue,
both the nuclear ($\sim 10$\,pc scale)
starburst and the AGN activity commonly needs efficient gas inflow to the circumnuclear 
and nuclear regions. Therefore, it is expected that intense star formation events around 
the nucleus of an AGN-hosting galaxy will provide us useful hints to understand the
triggering mechanism of AGNs. This issue is also important when we investigate 
the coevolution between galaxies and super massive black holes (SMBHs);
i.e., the positive correlation between the spheroidal and SMBH masses in galaxies
\citep{KH13, HB14}.\par

Since NGC\,1068 has intense circumnuclear star forming regions around its AGN,
it also provides us an important laboratory for this issue.
It has been suggested that NGC\,1068 has two stellar populations in the 100\,pc-scale 
circumnuclear region; around the nucleus \citep{SB12};
one is the relatively young stellar population with  an age of 300\,Myr
extending over the 100-pc scale circumnuclear region,
and the second one is the ring-like  structure at $\approx$ 100 pc from the nucleus with an
age of 30\,Myr. Since the inner 35 pc region is dominated by an old stellar population
with an age of $>$ 2\,Gyr, it is suggested that the two episodic intense star formation events
occurred in the circumnuclear region of NGC\,1068 although their origins have not yet been understood.
At the western part in the ring, molecular Hydrogen emission, H$_2$ S(1), is detected 
with a shell-like structure \citep{Sch00, Vale12}. 
Since this emission often probes the shock-heated gas, 
either a super-bubble or an AGN feedback effect or both have been discussed as its origin to date
\citep{SB12, GB14a, GB14b},
In either case, a certain asymmetric perturbation could drive
the intense star formation event 30\,Myr ago. \par

If there is a certain physical relationship between circumnuclear and nuclear
star formation events and the triggering AGN, it is intriguing to investigate 
the star formation activity in much inner region in NGC\,1068.
For this purpose,
it is essential to attain high spatial resolution down to pc-scale
both at dust continuum emission and thermal molecular lines,
which allow us to diagnose not only gas kinematics but also gas physics.
In this context, Atacama Large Millimeter/Submillimeter Array (ALMA) 
has been extensively used to study
atomic and molecular gas and dust properties of NGC\,1068 in detail
\citep{GB14a, GB14b, GB16, Ima16, Izu16}.\par

Among these brand-new ALMA observations, 
we emphasize potential importance of the newly detected 689\,GHz continuum source 
located close 
to the central engine of NGC\,1068 observed by \citet{GB14a}
(project-ID: $\#$2011.0.00083.S), 
although it was not identified as an independent source
by the authors (see their Figure 3). 
In addition, the continuum source has not been separately 
identified as an object in their CO\,(6--5) map either.
\citet{GB14a} interpreted that the molecular gas associated with the continuum source
represents a portion of the circumnuclear region rather than an independent source; 
see their Figure 4c.
Taking account of the proximity to the nucleus [the nuclear radio component S1 \citep{G04}], 
we consider that this source must be playing an import role to form
the observed complicated properties of the nuclear region of NGC\,1068.
In order to address the nature of the continuum source
and its role in the dynamics of the nuclear region,
we analyzed their ALMA data.\par

\section{Data}
\label{s:dt}
The ALMA data analyzed here were originally taken
by Garcia-Burillo et al.; see details of their observations in \citet{GB14a}.
We retrieved their image data from the data archive 
system of the Japanese Virtual Observatory\footnote{Japanese 
Virtual Observatory (JVO) is operated by 
National Astronomical Observatory of Japan (NAOJ).}.
We obtained the data set whose IDs are \texttt{ALMA01001360}
(the original file name of \texttt{NGC1068.B9.spw0.avg33chan.fits}) 
for the ALMA Band 9 CO\,(6--5) line image cube data, and 
\texttt{ALMA01001362} (\texttt{NGC1068.B9.continuum.fits}) 
for the Band 9 689\,GHz 
continuum image, which was obtained from concatenating 
four 1.875\,GHz bandwidth spectral windows by \citet{GB14a}.\par

We used the task \texttt{imhead} in the CASA package to 
set rest-frequency ($\nu_\mathrm{rest}$)
of the CO transition to be 
$\nu_\mathrm{rest} =\,691.473076$\,GHz. 
To increase signal-to-noise (S/N) ratio of the 
line emission, we smoothed the line data along the 
velocity axis every 2 channels using task \texttt{imrebin} by
keeping the original frame of ``LSRK" for the velocity axis. 
The resultant CO data have 58 channels with a resolution of 13.97 \kms. 
After completing this minimum data processing, 
we exported the CASA-formatted data into FITS files, 
and imported them into GILDAS package for scientific analysis.\par

Using our own scripts running on GILDAS package used in 
previous works, e.g., \citet{RSF14}, 
we shifted the origins 
of all the images to that of 
the previously known AGN; the nuclear radio component S1 \citep{G04}. 
Subsequently we evaluate RMS noise levels of the images 
by calculating statistics for arbitrary selected emission-free area 
in the 3-dimensional cube data.
Iterating such analyses by changing areas,
we found that RMS noise levels calculated in each velocity channel was fairly 
uniform with uncertainty of 28\%. We end up with the mean of the image noise levels of 
1\sgm\ $=$ 16.8 mJy \pbeam\ 
in specific intensity
per 14 \kms\ resolution for the CO\,(6--5) line, 
and 1.67 mJy \pbeam\ for the 689\,GHz continuum image.
Both the line and continuum images have the the same pixel size of \timeform{0.05"}.
The synthesized beam size of the images which we retrieved from the archive
(\timeform{0.33"}$\times$\timeform{0.22"} in FWHM at PA$=$\timeform{81D})
slightly differs from that in \citet{GB14a}
(\timeform{0.4"}$\times$\timeform{0.2"} at PA$=$\timeform{50D}).
We consider that such a difference would be caused 
by those in the visibility data flagging and parameters used
when Fourier-transformed into the image plane.\par

\section{Results}
\label{s:res}

\subsection{The Nuclear 689\,GHz Continuum Peak}
\label{ss:ContPeak}

Here we focus our attention
 on a nuclear 689\,GHz continuum peak
close to the nucleus.
In order to show the presence and the location of this
continuum  peak in the nucleus region of NGC\,1068, 
we present Figure \ref{fig:guidemap} where
overall morphology is shown by the optical image [panel (a)] whereas
the complicated morphology by the submm ALMA images [panels (b) and (c)]. 
We stress that the complexity of the central region is clearly recognized in
both (b) the velocity centroid map, which is produced from the CO\,(6--5) line, 
and (c) 689\,GHz continuum map.
Here, the velocity centroid map in unit of \kms\ is 
obtained as an intensity-weighted mean velocity map through dividing 
the first order momentum map by the zeroth order one. These moment 
maps are calculated by using the data shown in Figure \ref{fig:chmaps} 
with the task \texttt{moments}
over velocity range of 
1020 \kms\ $< v_\mathrm{sys}$(LSR)$<1240$ \kms. 
This velocity range is selected by comparing the velocity channel maps 
(Figure \ref{fig:chmaps}) and the spectrum (Figure \ref{fig:sp}). \par

Figure \ref{fig:guidemap}c presents the spatial distribution of 689\,GHz emission 
in the central region of the galaxy.
Comparing Figure \ref{fig:guidemap}c and the lower panel of Figure 3 in \citet{GB14a},
one immediately notices that there exists a local peak of the continuum
emission, but its position does not coincident with that of the known AGN, S1.
Although this continuum peak is readily recognized in Figure 3 of
\citet{GB14a}, these authors did not identify it as a distinct object
and any discussion was not given in their paper.\par

This 689\,GHz continuum local maximum has
the peak intensity of $I_{\tiny 689\mathrm{GHz}} =$ 16.2 mJy \pbeam, 
corresponding to 0.60\,K in 
the mean brightness temperature over the synthesized beam, 
$\langle T_\mathrm{sb}\rangle$.
We obtained its flux density of $S_{\tiny 689\mathrm{GHz}} =$ 9.5\,mJy 
integrated over the beam
centered on the peak.
This peak is located at
RA\,(J2000) $= 2^\mathrm{h} 42^\mathrm{m} 40^\mathrm{s}.714$ and 
DEC\,(J2000) $=$\timeform{-0D0'47.79"},
which is $\simeq$ \timeform{0.18"} NNE from
the position of the nuclear component S1 
identified by the 8.4\,GHz continuum imaging \citep{G04} at
RA\,(J2000) $= 2^\mathrm{h} 42^\mathrm{m} 40^\mathrm{s}.70912$ and
DEC\,(J2000) $=$\timeform{-0D0'47.9449"} 
which is adopted in \citet{GB16} (see caption of their Figure 1).
The position of S1 reported in \citet{G04} was obtained through astrometry 
between the 8.4\,GHz continuum image taken by VLBA and that of 
the \wat\ masers observed by
VLA, yielding absolute position accuracy of $\sim 1.2$ mas [see Figure 7 in \citet{G04}].
On the other hand, it is not trivial to evaluate absolute position accuracy 
of the 689\,GHz continuum peak, 
which should be primarily determined by 
accuracies of the baseline vectors, 
angular separation(s) to the calibrator(s), and 
their absolute position accuracies.
We therefore arbitrary assume the widely accepted idea that 
absolute position accuracy of a point source imaged by a connected-type interferometer 
is typically better than a fraction of its
synthesized beam size.
Namely we employ an absolute position accuracy to be
$\sqrt{\timeform{0.328"}\times \timeform{0.215"}}/5\sim \timeform{0.05"}$
as a fiducial value. 
Notice that the accuracy is
identical to the pixel field of view (\S\ref{s:dt}).
Taking all the above into account, the angular separation of \timeform{0.18"}
between the S1 and the 689\,GHz continuum peak is believed to be real,
as clearly recognized in Figure 3 of \citet{GB14a}.
Last, the angular separation corresponds 
to the projected separation of $\simeq$ 14 pc.\par

\subsection{Molecular Gas Associated with the 689\,GHz Continuum Peak}
\label{ss:ResMolGas}

In order to elucidate the origin of  the 689\,GHz continuum source,
we investigate molecular gas properties associated with this source.\par

First, we compare velocity channel maps of the CO (6--5) line emission 
by overlaying on the continuum emission map in Figure \ref{fig:chmaps}.
We note that the CO emission around the continuum peak appears to be 
contaminated with the gas associated 
with the circumnuclear ring \citep{Sch00}, 
as seen towards the top-left corner in each panel. \par

Second, we present CO\,(6--5) line spectral profile towards the continuum peak 
after subtracting the continuum emission
(Figure \ref{fig:sp}).
The spectrum was made by integrating  the CO\,(6--5) emission inside the dotted-ellipse 
shown in Figure \ref{fig:chmaps}. 
The ellipse, i.e., the adopted aperture, is centered on the 689\,GHz continuum peak 
and the area is identical to that of the synthesized beam size.
The systemic velocity of the entire galaxy, \Vsys(LSR)\,$=$\,1126 \kms, 
falls between the blue and green bars in the spectrum, 
yielding asymmetry of the blue- and red components with respect 
to the systemic velocity, \Vsys. 
We thus believe that the local gas associated with the continuum is decoupled from 
the galaxy-wide motion of the gas. 
Therefore we arbitrary assume that the green-coded component seen in
the velocity range of
$v_\mathrm{green}$(LSR)\,= 1140-- 1160 \kms\ 
represents the bulk motion of the local gas. 
This assumption would not be affected by the results from the
higher angular resolution new observations by \citet{GB16} because
the source of our interests appears to be resolved out
by the extended array configuration observations.
The blueshifted component is seen in 
$v_\mathrm{blue}$(LSR)\,= 1050 -- 1120 \kms, 
whereas the red one at $v_\mathrm{red}$(LSR)\,= 1200 -- 1230 \kms. 
Namely, both the blue- and redshifted components have almost same velocity shifts 
with respect to that of the bulk gas, i.e., 
$\delta v = | v_\mathrm{blue, red} - v_\mathrm{green}\mathrm{(LSR)}| \simeq$ 65 \kms.\par

Third, in Figure \ref{fig:pv}, we show the position-velocity (PV) maps where 
we adopted the line with PA $=$\timeform{120D} passing at the 689\,GHz 
continuum peak as a slicing line 
(the solid line in Figure \ref{fig:guidemap}b).
The direction of the line is perpendicular to the line 
connecting the 689\,GHz continuum peak with the nucleus, S1 
(PA$=$\timeform{30D}). 
We also point out that the 689\,GHz continuum is elongated towards 
the north-northeast (PA$\sim$\timeform{30D}); see Figure \ref{fig:guidemap}b.
Despite inadequate spatial resolution, 
the PV diagram in the panel b demonstrates that
multiple velocity components of the gas coexists within the
compact region which cannot be resolved by the beam size of 
the data analyzed in this work.\par

Last, we do not completely rule out an alternative hypothesis 
that an opaque ``static" single-velocity component gas is
responsible for the multiple velocity features 
in Figures \ref{fig:sp} and \ref{fig:pv}.
In this interpretation, the spectral profile is considered as self-absorption of the line
because of high optical depth 
($\tau_\mathrm{CO6-5}\sim 15$; described in \S\ref{ss:SFandGproperty}).
However, this single gas hypothesis has a caveat that
one should observe a double-peaked spectral profile
whose absorption dip appears around
the LSR-velocity of the static bulk gas.
Contrary to this expectation, 
we detected the weak emission labeled with green bar (Figure \ref{fig:sp}).
We therefore stick on the inference from Figure \ref{fig:pv} that there exist
multiple components
of the gas having different velocities along the line of sight.

\section{Analysis}
\subsection{Dynamical Properties}
\label{ss:DP}
Subsequent questions are how compact the gas cloud is and 
what is the origin of the multiple velocity components.
An estimate of the size may be obtained 
from effective radius (\Reff) of the beam size, 
i.e., the geometrical mean of the major and minor axes of elliptical beam
of \Reff\ $=$\,21\,pc. 
Although the beam size does not suffice to resolve the gas ``condensation",
we attempt to give a better constraint on the radius as follows.
At each velocity channel, we search for a peak pixel within the aperture 
where we made the spectrum.
Figure \ref{fig:pixpos_summary} compares the peak pixel positions
obtained from each velocity channel shown in Figure \ref{fig:chmaps}.
To produce the figure, 
we limited to plot the peak pixel positions which have S/N-ratio of higher than 4. 
Assessing scatter of their positions, 
we obtained a stronger constraint that the spatial extent of gas 
is at most 2\Reff $\sim$\timeform{0.07"}, 
corresponding to 2\Reff$\sim$5\,pc 
in effective diameter (see Figure \ref{fig:pixpos_summary}). 
Here we excluded the $v_\mathrm{sys}$(LSR)\,$= 1122$ \kms\ 
component which seems to represent a local maxima of the gas
contaminated with that associated with the circumnuclear ring rather than
the gas of our interests.\par

Taking account of both the symmetry of velocity ranges where signals were detected and 
the spatio-velocity structure recognized in Figure \ref{fig:pv}b, 
one may consider that the blue- and redshifted gas 
are associated with a rotating structure around the central object.
If we adopt a rotating radius $r=$ \Reff $\sim$3\,pc, 
an upper limit of the enclosed mass is calculated to be
$M_\mathrm{rot} \leq 3 \times 10^6 \Msun\
\left(\frac{r}{3\,\mathrm{pc}}\right)  
\left(\frac{v_\mathrm{rot}}{65\,\mathrm{km\,s^{-1}}}\right)^2$
where we set $v_\mathrm{rot}\,= \delta v$.
However, we argue that the gas is not in equilibrium by a pure rotation.
This is because its specific angular momentum of
$\log_{10}{\mathrm J}\,\equiv\,\log_{10} (0.4r^2\Omega)\sim 25$
is significantly higher than that expected from 
the correlation between $\log_{10}{\mathrm J}$ and $r$ \citep{Bodenheimer95}
where $\Omega$ is angular velocity of 
$\Omega = v_\mathrm{rot}/r \sim
65\,\mathrm{km\,s^{-1}} / 3 \,\mathrm{pc} =  7\times 10^{-13}$ s$^{-1}$ for this object.
Note that a typical GMC has $\Omega$ of the order of 
$10^{-15}$ s$^{-1}$ \citep{Bodenheimer95}.
We therefore return to the most na\'ive hypothesis:
there are multiple components of gas having different velocities along the
line of sight.\par

\subsection{Star Formation and Gas Properties}
\label{ss:SFandGproperty}

Another clue to shed light on the nature of the gas condensation is
obtained from analysis of the continuum flux.
Following the spectral energy distribution (SED) analysis in \citet{GB14a}, we attempted to explain the
observed 689\,GHz continuum flux by thermal emission from dust 
grains, which can be approximated by a single temperature grey body emission.
Adopting a range for dust temperature ($T_\mathrm{d}$) of 50 --70\,K,
frequency-index of emissivity of dusts ($\beta$) of 1.7 \citep{Klaas01}, 
dust mass absorption coefficient ($\kappa_0$) at reference 
frequency ($\nu_0=\,231$ GHz) of 0.005 cm g$^{-1}$ \citep{Preibisch93, Andre96}
and the \Reff\ value, 
we found that the gas plus dust mass of $M_\mathrm{gas} = (5\pm 3)\times 10^5$ \Msun\
is required to reproduce the observed $S_\mathrm{690\,GHz}$ value.
For a simplicity  of the analysis, we kept the hypothesis that the thermal emission
from dust grains can be approximated as if it is emanated from a single component gas
[e.g., \citet{Klaas01}],
regardless of the possible multiple ones (\S\ref{ss:ResMolGas}).
Notice that the adopted $\kappa_0$ value is a typical one for
interstellar medium whose spectral energy distribution often shows $\beta\sim 2.0$ 
[e.g., \citet{Beckwith00}], $\sim 1.8$ \citep{Klaas01}, and 1.78 \citep{PC11b}.
It should be also noticed that the above $\kappa_0$ value is not for dusts alone,
but for whole the interstellar medium, therefore so-called dust-to-mass ratio is not
needed to be multiplied.\par

The inferred $M_\mathrm{gas}$ and \Reff\ yield mean molecular hydrogen
number density of $n_\mathrm{H_2}\sim 1\times 10^5$ cm$^{-3}$,
which is comparable to the critical density of CO\,(6--5) transition.
Because of such a high density, we refer to the continuum source as 
``massive dense gas cloud", which would be a scaled-up version
of the galactic high-mass star-forming hot molecular cores (HMCs)
[see e.g., \citet{Kurtz00, Beltran05, RSF11}].\par

Furthermore, the $M_\mathrm{gas}$ value
leads the mean column density of $\langle N_\mathrm{H_2}\rangle$ of
$(7\pm 2)\times 10^{23}$ cm$^{-2}$, yielding
mean optical depth of $\langle \tau_\mathrm{689\,GHz}\rangle$ of the
order of 0.01 -- 0.1. 
Since bolometric luminosity of optically thin dust emission is given by
$L_\mathrm{bol}\, = \frac{8\pi h}{c^2}\,
\frac{\kappa_0 M_\mathrm{gas} }{\nu_0^\beta}\,
\left(\frac{kT_\mathrm{d}}{h}\right)^{\beta +4}\,
\zeta(\beta +4)\,\Gamma(\beta +4)$
where $\zeta$ denotes Riemann's zeta function
and $\Gamma$ gamma function, we estimate that $L_\mathrm{bol}$ 
would range (0.4 -- 4)$\times 10^8$ \Lsun.
If the widely accepted conversion factor between infrared luminosity 
($L_\mathrm{FIR}$) and
star formation rate (SFR) given by Eq.(4) in \citet{K98} can be applied to the
gas, we calculated SFR to be $\sim$  0.1 \Msun\ yr$^{-1}$
with another assumption of $L_\mathrm{bol}\approx L_\mathrm{FIR}$.
Hereafter we summarize derived properties in Table \ref{tbl:Ps}. \par

Given the resultant $\langle N_\mathrm{H_2}\rangle$ value 
and a fractional abundance of 
[$^{12}$C$^{16}$O]/[H$_2$] $\sim 10^{-4}$ \citep{Dickman78}, and assuming 
that the gas and dust are well-coupled, i.e., 
gas temperature ($T_\mathrm{gas}$) is represented by the $T_\mathrm{d}$ of \,50--70\,K, 
we obtained radiation temperature ($T_\mathrm{R}$) for the CO (6--5) transition to be 35--52\,K
and optical depth of the line of $\tau_\mathrm{\tiny CO6-5}\sim 15$
using non-LTE radiative transfer code of \texttt{RADEX} \citep{vdTak07}
for the most intense blueshifted component (Figure \ref{fig:sp}).
In this calculations, we keep the practical approximation of the single-gas hypothesis, 
and adopted that
the blue component has velocity width ($\Delta v_\mathrm{FWHM}$) of $\sim 30$ \kms\ in FWHM
and the volume density estimated above.
We also measured that it has the peak flux density of $S_\nu\sim$ 70\,mJy (Figure \ref{fig:sp}),
which corresponds to $\langle T_\mathrm{SB}\rangle \simeq 2.5$\,K.
The ratio of $\langle T_\mathrm{SB}\rangle / T_\mathrm{R}\sim 0.05$ gives an estimate
of beam-filling factor ($f_\mathrm{b}$) for an optical thick line.
From the definition, $f_\mathrm{b}$ is calculated by $f_\mathrm{b}\simeq\left(r/21\,\mathrm{pc}\right)^2$
where $r$ is the desired radius of the gas,
we obtain $2r\sim 10$\,pc by solving 
$\langle T_\mathrm{SB}\rangle / T_\mathrm{R}\sim \left(r/21\,\mathrm{pc}\right)^2$.
Repeating the same analysis for the red, 
we obtained $2r\sim 8$\,pc.
Regardless of such a robust assessment, 
we confirmed that the inferred $2r\sim 10$\,pc has a reasonable consistency 
with the estimate of 2\Reff$\sim$5\,pc which is independently obtained from Figure \ref{fig:pixpos_summary} (\S\ref{ss:DP}).
We confirmed that the number density of the gas, $n_\mathrm{H_2}$, 
inferred in the second paragraph of this subsection remains the same 
within a factor of 2--3. even if we calculate $n_\mathrm{H_2}$ 
with the revised \Reff.\par

The \Reff\ and a 3D velocity dispersion of
$\sigma_\mathrm{3D}=\frac{\Delta v_\mathrm{FWHM}}{\sqrt{8\ln 2}}\sqrt{3}$
yield an ``effective" virial mass
which includes both thermal and non-thermal contributions to support the gas
against self-gravity,
of $M_\mathrm{vir}\sim 6\times 10^5$ \Msun .
Because of $M_\mathrm{gas}\approx M_\mathrm{vir}$, the gas must be
on the verge of star formation.
Moreover it is fairly reasonable to conclude that formation of a star or a star cluster
has already commenced in the gas because such a gas 
would gravitationally collapse within 
a few times of free-fall time of 
$\tau_\mathrm{ff}\sim 10^5\,\mathrm{yrs}
\left(\frac{n_\mathrm{H_2}}{10^5\mathrm{cm}^{-3}}\right)^{-1/2}$
or within a dissipation time scale of turbulence of
$\tau_\mathrm{disp}\sim 4\times 10^5 \,\mathrm{yrs}
\left(\frac{2R_\mathrm{eff}}{10 \mathrm{pc}}\right)
\left(\frac{\sigma_\mathrm{3D}}{10 \mathrm{pc}}\right)^{-1}$.
Notice that our SED analysis adopts an implicit assumption that 
the 689\,GHz continuum emission is purely due to reprocessed thermal emission 
from dust grains heated by  internal sources.
On the basis of the $M_\mathrm{gas}$ value, 
which is comparable to those of galactic giant molecular clouds, and 
the fact of $M_\mathrm{gas}\approx M_\mathrm{vir}$, 
we further discuss that the putative heating source which is deeply embedded in
the ``massive dense gas cloud" is a ``proto starcluster".\par

Assuming a typical star formation efficiency (SFE), which is defined by 
$\varepsilon \equiv M_\ast/(M_\mathrm{gas}+M_\ast)$, of a few to 10\%
measured in the galactic star-forming clouds \citep{KE12},
the gas may form (or may be forming)
a ``proto starcluster" with a total
stellar mass ($M_\ast$) of on the order of 
$M_\ast \sim  \varepsilon M_\mathrm{gas} \sim$  a few $\times 10^4 \Msun$
(Table \ref{tbl:Ps}).\par

\section{Discussion}
\label{s:dis}

\subsection{Nature of the Massive Dense Gas Cloud}
\label{ss:nature}

It is interesting to compare physical properties of 
``the massive dense gas cloud"
in the vicinity of the nucleus of NGC\,1068 with those in our Galaxy 
and other galaxies because it provides us with a crucial hint on its origin.
For this purpose, we present a plot of SFR vs. $M_\mathrm{gas}$ in 
Figure \ref{fig:SFR}a which is taken from \citet{Lada12}.
Clearly, the estimated SFR with an order of 0.1 \Msun\ yr$^{-1}$
and the gas mass of $2 \times 10^5$ \Msun\ (Table \ref{tbl:Ps}) make 
the ``massive dense gas cloud" unique in the diagram
located at almost middle 
between the galactic star-forming clouds and in extra-galaxies.
Moreover we point out that dense gas fraction of the cloud, $f_\mathrm{dg}$,
is almost 100\%, as expected from the high $n_\mathrm{H_2}$ value described above.
The $f_\mathrm{dg}$ is comparable not only 
to those of intense star-forming gas clouds observed in 
ultraluminous infrared galaxies (ULIRGs) such as Arp\,220, but also
those of active star-forming gas clouds in our Galaxy.
\par

Next, we investigate properties of the gas with 
the relationship between the SFR surface density,
 $\dot{\Sigma}_\ast$  in unit of \Msun\ yr$^{-1}$ kpc$^{-2}$, 
and the surface gas density,
$\Sigma_\mathrm{g}$ in \Msun\ pc$^{-2}$ (Figure \ref{fig:SFR}b),
which is originally produced by \citet{KM05}. 
The dashed line in the plot indicates the best-fit curve for the observed data \citep{K98}, and
the solid one is the analytical prediction from the turbulence-regulated star formation model
by \citet{KM05} which adopts the following three assumptions; 
{\it (1) star formation occurs in molecular clouds 
that are in supersonic turbulent state,
(2) the density distribution within these clouds is lognormal},
and  {\it (3) stars form in any subregion of a cloud that is so overdense that its
gravitational potential energy exceeds the energy in turbulent motions.}
We argue that the ``massive dense gas cloud" satisfies the
three assumptions.
First, the results of 
$M_\mathrm{gas}\gtrsim M_\mathrm{vir}\gg M_\mathrm{vir}^\mathrm{thm}$ 
indicates the commence of star formation, which is
suggested presumably to be under a turbulent status.
Here $M_\mathrm{vir}^\mathrm{thm}$ is a virial mass that can be supported by sole thermal motion of
$M_\mathrm{vir}^\mathrm{thm}\sim 
\frac{k T_\mathrm{gas}}{\mu m_\mathrm{H}}
\frac{R_\mathrm{eff}}{G}
\sim \,250 \Msun $ for the 50--70\,K gas where
$\mu$ is a mean molecular weight (2.33 for [He] = 0.1 [H]).
This leads a ratio of
$M_\mathrm{vir}/M_\mathrm{vir}^\mathrm{thm}\propto 
\left(\sigma_\mathrm{nth}/\sigma_\mathrm{thm}\right)^2
=$ (Mach number)$^2$ to be an order of $10^3$ (Table \ref{tbl:Ps}), 
suggestive of highly turbulent status.
Here $\sigma_\mathrm{nth}$ and $\sigma_\mathrm{thm}$ are 
non-thermal and thermal velocity dispersions, respectively.
The second assumption is not readily proven without
a detailed analysis e.g., Figure 12a in \citet{RSF14}, but
the above $M_\mathrm{vir}\gg M_\mathrm{vir}^\mathrm{thm}$
will allow us to hypothesize it.
The third one is supported by the multiple velocity components of the gas (\S\ref{ss:DP}).\par
Although we need to have higher resolution observations with higher sensitivity 
to assess physical properties of the gas on more firm ground, 
its star formation activity may be explained in terms of the turbulence-regulated
star formation scenario.
Last, it is interesting that 
the location of the massive dense gas in the $\Sigma_\mathrm{g}-\dot{\Sigma}_\ast$ plane
is close to those of SSCs in ULIRGs.\par

\subsection{Origin of the Massive Dense Gas Cloud}
\label{ss:origin}

We investigated physical and star-formation properties of the 
``massive dense gas cloud" in NGC\,1068.
Now, we briefly address how such a star cluster is formed 
in the very nuclear region of the galaxy.
A straightforward interpretation is that the ``massive dense gas cloud"
was formed through a shock compression of clouds via 
cloud-cloud collision \citep{Habe92, Hasegawa94, IF13} in the nuclear region. 
If this is the case, a question arises as how such a cloud collision was induced 
in the very nuclear region
of NGC\,1068.\par

Here we remind again that  NGC\,1068 has two distinct star forming regions around the nucleus;
one is the so-called circumnuclear star forming region whose star formation activity has an age of 300\,Myr,
and the second one is the ring-like structure at $\approx$100\,pc from the nucleus with an
age of 30\,Myr \citep{SB96}. 
This means that NGC\,1068 experienced a couple of episodic star formation events
in their circumnuclear regions.
If we assume that NGC\,1068 experienced a minor merger in the past,
recurrent star formation events induced by cloud-cloud collision can be naturally understood 
because they are induced by the orbital sinking motion of a satellite galaxy to be merged \citep{MH94, TW96}.
Before adopting such interpretation, a caution must be used because recent 3D magnetohydrodynamics simulations 
pointed out that a galaxy itself can form such massive dense gas clouds by means of
collision of filamentary clouds threaded by magnetic fields \citep{IF13}
or by multiple compressions \citep{SI15} without merging of galaxies.\par

However, in the case of NGC\,1068, 
some observational properties suggest a past minor merger,
although any disturbed structures cannot be recognized around the galaxy
(see Figure \ref{fig:guidemap}a).
First, kpc-scale narrow line regions are distributed 
along an axis which is far from the rotational axis of the galactic disk \citep{Cecil90, Cecil02}. 
Second, the molecular torus ($\sim$0.1\,--\,1\,pc scale) probed 
by \wat\ masers is observed as almost the edge-on geometry
\citep{Greenhill96, G96_H2O, G01}, 
whereas the overall galactic disk is observed to be a nearly face-on geometry\footnote{
The observed optical minor-to-major axis ratio of 0.85 nominally gives 
the viewing angle toward the galactic disk of NGC\,1068 to be 
\timeform{32D} \citep{dV91}.
}.
Third, circumnuclear molecular gas clouds ($\sim$100\,pc scale) also show 
highly asymmetric structures \citep{GB14a}. 
All these lines of evidence can be interpreted as that 
the dynamics of the nuclear region would be decoupled from 
that of the entire galactic disk.
These characteristic properties would not be readily explained 
if the gas inflow in NGC\,1068 were due to gradual angular momentum loss
driven by such as spiral arms and a bar structure in the galactic disk.
Therefore, the gas fueling driven by minor-merger
seems to be the most natural
mechanism for the case of NGC\,1068
[see for a review of \citet{YT99}].
It should be noticed that a minor merger would occur by taking an inclined orbit
with respect to the galactic disk of a host galaxy, 
making both circumnuclear
and nuclear structures decoupled from the dynamics of the galactic disk.
It is also reminded that the orbital period becomes shorter as the separation
between the satellite and the host galaxy becomes smaller.
Namely, the satellite galaxy is anticipated to
interact or collide with the galactic disk more often over a certain period 
whose time scale becomes shorter as the merger stage proceeds
\citep{MH94, TW96}. 
This explains the observed nature of episodic star formation events
in NGC\,1068. 
Considering the well-defined overall symmetric morphology of the outer disk,
we propose a picture that NGC\,1068 is experiencing the final stage of a minor merger.
In this context, we argue that 
the newly found ``massive dense gas cloud" having SFR of the order of 0.1 \Msun\ yr$^{-1}$
may be formed by past gas collision(s) between/among nuclear gas clouds in
the putative minor merger event.\par

Another merit of the minor merger scenario is that star clusters 
can be formed in the central region of a merger remnant \citep{MH94, TW96}.
From an observational ground, 
massive star clusters known as super star clusters (SSCs) often form
in the interacting regions of major mergers such as in luminous infrared galaxies (LIRGs)
[e.g., \citet{Whitmore93, Mulia16}] and ULIRGs [e.g., \citet{Shaya94, Shioya01}].
In the case of Arp\,220, some SSCs in the central region tend to be massive 
(e.g., $M_\ast \sim 10^8 \Msun$) than those located in the circumnuclear zone 
[e.g., $M_\ast\leq 10^6 \Msun$: \citet{Shioya01}].
On the other hand, in the case of NGC\,3256, one of luminous infrared galaxies (LIRGs),
the typical mass of the nuclear star clusters is also $M_\ast \leq 10^6$ \Msun\citep{Mulia16}.
On the other hand, as for minor mergers, such observations have not yet been made to date.
In general, it is difficult to identify galaxies in a late phase of a minor merger
because tidal features in the outer part of the galaxy were easily smeared out
after several rotations of the galactic disk [e.g., \citet{Khan12}].
Clearly, it requires to conduct systematic surveys for minor mergers
in Seyfert galaxies, and then carry out high-resolution optical imaging 
to search for nuclear star clusters in these system.\par

It is important to note that the ``massive dense gas cloud" appears to be associated with 
the nucleus of NGC\,1068 at a projected separation of 14\,pc.
Since the nucleus, i.e., a SMBH, has a mass of  
$M_\mathrm{SMBH} \simeq (8.4\pm 0.4) \times 10^6$ \Msun \citep{KH13}, 
the SMBH system with accompanying the ``massive dense gas cloud"
is expected to behave as a binary with this SMBH, yielding
an asymmetric gravitational potential. 
It is possible that this explains the complicated observational properties 
in the nuclear region of NGC\,1068.\par
It is also worthwhile to note that the \wat\ maser disk (or ring) around 
the SMBH at the nuclear radio component S1 in NGC\,1068 does not exhibit pure Keplerian 
rotation \citep{Greenhill96, G96_H2O, G01, MT97}, 
whereas that of NGC\,4258 is explained almost perfectly by 
a Keplerian rotation \citep{Miyoshi95}. 
As also discussed in \citep{GB16}, the observed non-Keplerian motion could be 
a signature the so-called Papaloizou-Pringle instability \citep{PP84},
although it still remains possible that the non-Keplerian rotation  
may be attributed to the dynamical interaction with the star cluster.\par

AGNs are thought to be powered 
by the gravitational energy release through the gas accretion  (i.e., the gas fueling)
onto a SMBH resided  in the nucleus of galaxies \citep{Rees84}. 
Among several physical mechanisms for such efficient gas fueling, 
galaxy major mergers appear to be the most efficient mechanism to explain 
the triggering AGN phenomena \citep{Sanders96, Hopkins08}.
If low-luminosity AGNs such as Seyfert galaxies can be powered by minor mergers
with a satellite galaxy, it is possible to have a unified triggering mechanism for
all types of AGNs \citep{YT99, YT13}. 
Such future studies will provide us with unique
opportunity to test our knowledge of star formation, 
which is established in the galactic ``quiescent" clouds,
in the extreme environments, 
such as in the nucleus of Seyfert galaxies and merging galaxies.\par

\section{Concluding Remarks}
To shed light on the nature of both circumnuclear and nuclear star formation 
in conjunction with the AGN activity, 
we analyzed ALMA archival data on both CO\,(6—5) line and
689\,GHz continuum emission towards 
the archetypical nearby Seyfert galaxy NGC\,1068 ($d$ =15.9\,Mpc). 
The ALMA data were originally taken
by Garcia-Burillo and colleagues [see details of their observations in \citet{GB14a}].
In this work, we focused on the 689\,GHz local continuum peak 
in the vicinity of the nucleus, located at 14\,pc (\timeform{0.18"}) 
NNE from the nucleus.
Although the continuum peak of our interests was 
already found in the analysis by \citet{GB14a}, 
no discussion was given in their paper. 
Since a near-nuclear gas condensation 
such as the newly identified ``massive dense gas cloud" is 
generally expected to physically affect the nuclear activity of a galaxy, 
we thoroughly investigated the physical properties of the source. 
Our findings can be summarized as follows.\par

\begin{enumerate}
\item The 689 GHz continuum flux gives a gas mass 
and bolometric luminosity (see Table \ref{tbl:Ps} for the values), 
allowing us to estimate to a SFR of  $\sim$ 0.1 \Msun\ yr$^{-1}$. 
We estimated size of the gas to be $\sim 10$\,pc
in diameter by means of two methods (\S\ref{ss:DP} and \S\ref{ss:SFandGproperty}).
Because both results have a reasonable consistency,
we obtained a mean H$_2$ number density of
$\sim 10^5$ cm$^{-3}$. Therefore, this continuum peak can be identified as
a ``massive dense gas cloud”.

\item  The gas density is high enough to form a ``proto starcluster" with
a total stellar mass of
$M_\ast\sim 10^4$ \Msun.
We argue that this gas cloud will evolve to a nuclear star cluster
around the nucleus of NGC\,1068.

\item  The gas cloud is identified as a missing link between
galactic and extragalactic gas clouds in the previously
known scaling relations of
[a] SFR vs. gas mass proposed by \citet{Lada12},
and [b] surface density of gas vs. SFR density by \citet{KM05}.
All the gaseous and star-formation properties 
(Table \ref{tbl:Ps} and Figure \ref{fig:SFR}) may be understood
in terms of the turbulence-regulated star formation scenario proposed by  \citet{KM05}.

\item  Since there are two stellar populations with the ages of 300\,Myr and 30\,Myr
in the 100 pc-scale circumnulear region, we discuss that 
NGC\,1068 has experienced at least three episodic star formation events
with a tendency that 
inner star-forming region is young in its age of star formation.
Given the evidence for the gas dynamics in the nuclear region, 
the nuclear region of NGC\,1068 is suggested 
to be decoupled from that of the entire galactic disk.
We propose that the gas inflow towards 
the nuclear region of the galaxy may be driven by a past minor merger.

\end{enumerate}

\begin{figure}[ht!]
\includegraphics[angle=0,scale=.54]{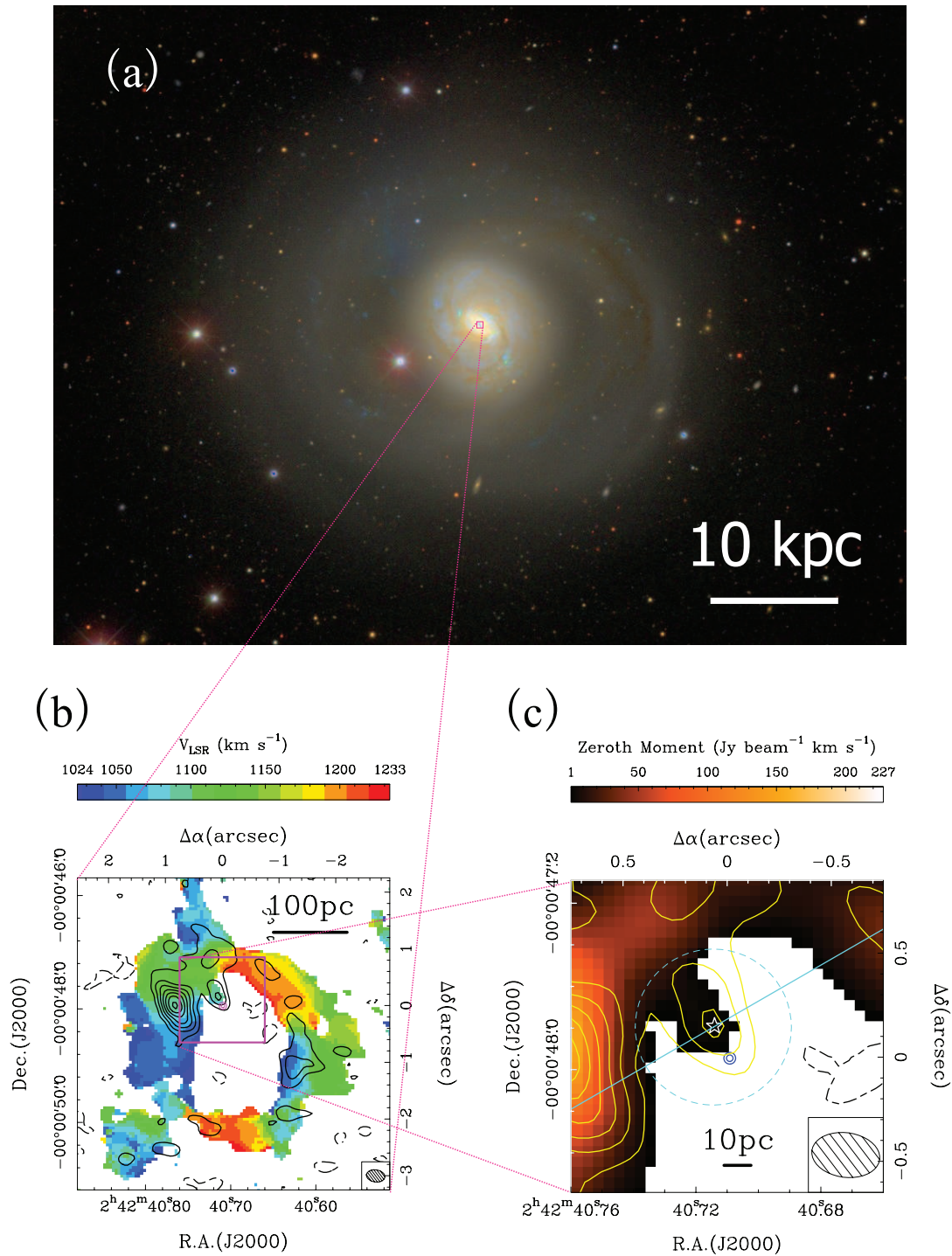}
\caption{
(a) An optical color image of NGC\,1068 based on the Slocan Digital Sky Survey 
data (courtesy of Michael R. Blanton). 
The image size is \timeform{14.22'}$\times$\timeform{12.11'}, 
and the north is up and east is left. The horizontal white bar at the bottom-right 
corner indicates a linear scale of 10\,kpc at $d$\,=\,15.9 Mpc.
(b) An overlay of the 689\,GHz continuum map (contours) on  
the velocity centroid map of the CO\,(6--5) emission (color) 
in the \timeform{7.5"} squared region centered at the previously known AGN, S1.
(c) The positions of the 689\,GHz peak (star) and the known AGN, S1, (double circles)
shown on the continuum map (yellow contour) and  
total integrated intensity map of the CO\,(6--5) emission (color).  
All the contour intervals are 3\sgm\ steps, 
starting from the 3\sgm\ level, 
while the dashed contour shows the $-3$\sgm\ level.
See the text in \S\ref{s:dt} for  the image noise level.
The light-blue dashed-circle with a radius of \timeform{0.45"} indicates the region 
where we performed our position-velocity (PV) diagram analysis 
shown in Figure \ref{fig:pv}a. 
The slicing axis for the PV diagrams is shown by the light-blue dashed-line. 
\label{fig:guidemap}}
\end{figure}

\begin{figure}[ht!]
\includegraphics[angle=0,scale=.48]{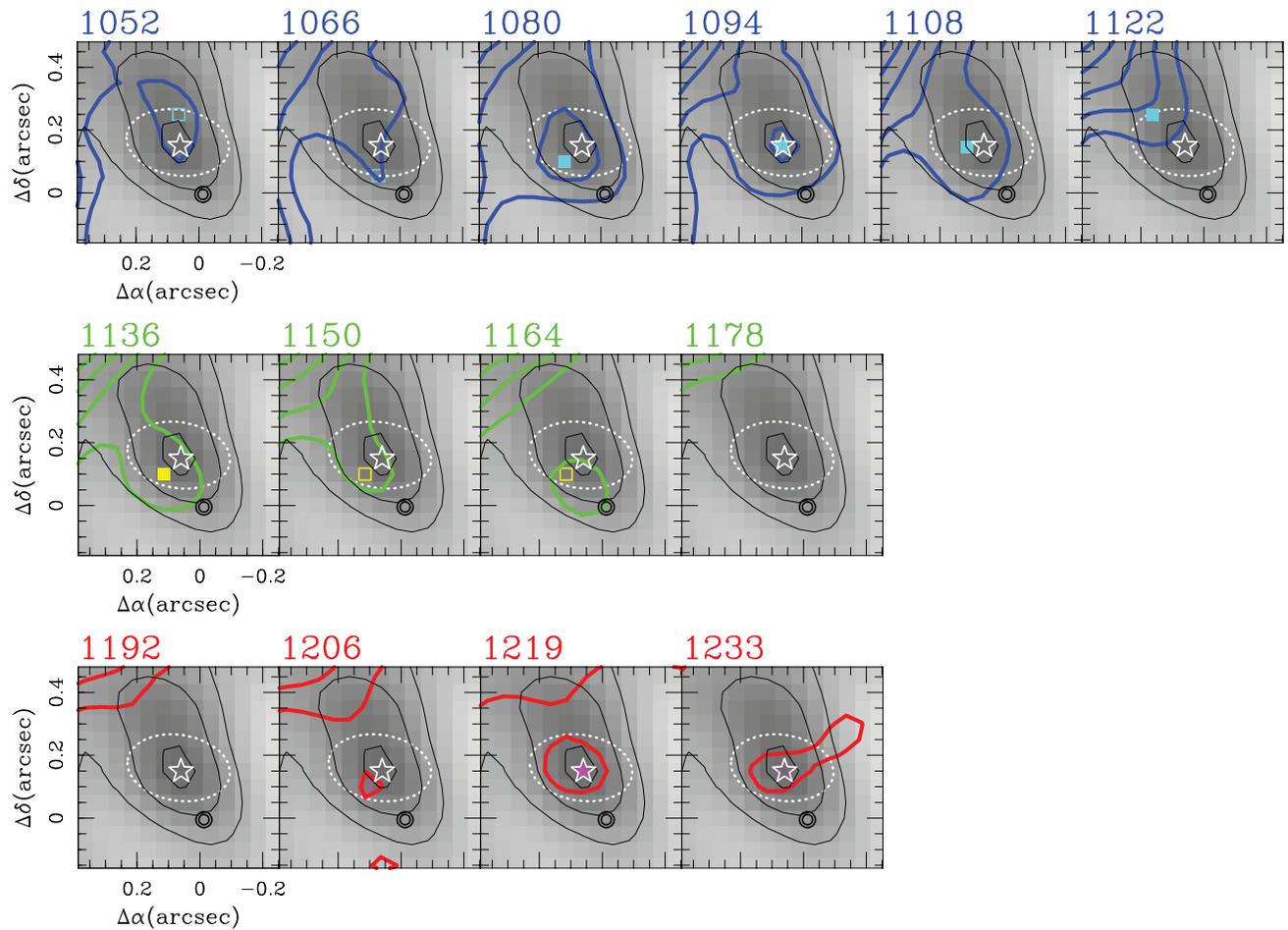}
\vspace{0truemm}
\caption{
Velocity channel maps of the  CO\,(6--5) emission (color contours) 
in  the  \timeform{0.5"} square nuclear region of NGC\,1068. 
The 689\,GHz continuum emission map is also shown in each panel
by grey scale plus thin contours.
The continuum peak and the nuclear radio source S1 are shown 
by white star and black double circles, respectively.
Color square in each panel indicates the peak pixel position 
of the CO\,(6--5) emission. All the filled squares have S/N $\geq$ 4 while
the open ones have 3$<$ S/N $<$4.
The dotted-ellipse in each panel indicates 
the aperture adopted for producing the CO spectrum in Figure \ref{fig:sp}. 
Note that the colors of contours correspond to those 
of the three vertical bars in Figures \ref{fig:sp} and \ref{fig:pv}b. 
Contour intervals are 3\sgm\ steps, starting from the 3\sgm\ level.
\label{fig:chmaps}}
\end{figure}

\begin{figure}[ht!]
\includegraphics[angle=0,scale=.4]{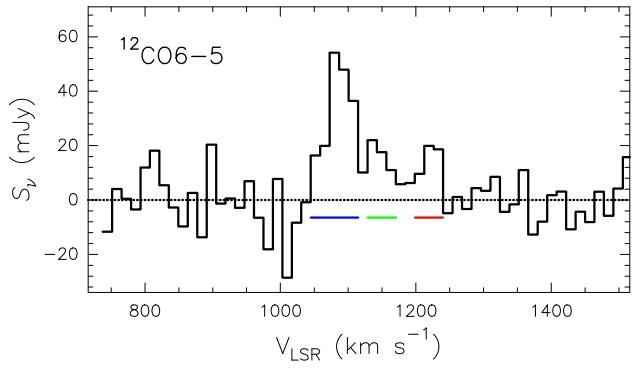}
\vspace{32truemm}
\caption{Interferometric spectrum of the CO\,(6--5) emission towards 
the peak position of the 689\,GHz continuum
in flux density, $S_\nu$, obtained by integrated the emission in 
$I_\nu$ over the beam solid angle.
The horizontal blue, green, and red bars 
indicate their velocity ranges where the detection of signals 
is assessed either in the spectrum or PV diagram (Figure \ref{fig:pv}b). 
The continuum emission with the mean $S_\nu$ of 15\,mJy was subtracted in
the spectrum shown in this figure.
Here the mean value was calculated over the line-emission free channels of
the spectrum, and 
the RMS noise level of the spectrum in $S_\nu$ is 
9.9 mJy with the velocity resolution of 14\,\kms.
\label{fig:sp}}
\end{figure}

\begin{figure}[hb!]
\includegraphics[angle=0,scale=.5]{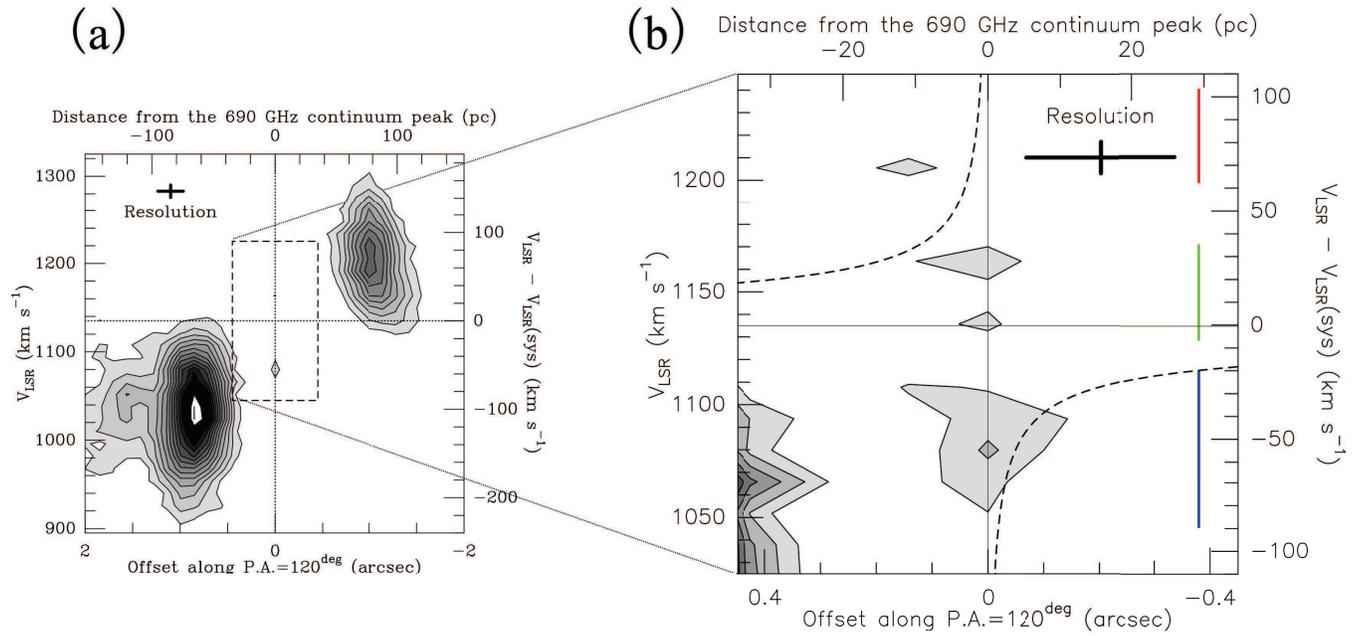}
\vspace{0truemm}
\caption{(a) PV diagram of the CO (6--5) emission towards the central region of  NGC\,1068,
produced with the velocity 
channel maps shown in Figure \ref{fig:chmaps} along the PA$=$\timeform{120D} line 
passing the 689\,GHz continuum peak (see the light-blue line in Figure\,\ref{fig:guidemap}c). 
The upper horizontal axis indicates position offset from the 689\,GHz continuum 
peak, and the right vertical axis
 does the velocity offset with respect to the systemic velocity of the galaxy, 
$v_\mathrm{sys}$(LSR) $= 1127$ \kms\ \citep{GB14a}. 
Contours are plotted by an interval of 3\sgm\ starting from the 3\sgm\ level.
The spatial resolution of \timeform{0.27"}, which corresponds to 21\,pc when projected 
along the slicing axis, and the velocity resolution are shown by the horizontal
and vertical thick-bars, respectively, at the top-left corner of the panel. 
The dashed rectangle region presents the area shown in panel (b). 
(b) Close-up view of the PV diagram towards the 689\,GHz continuum source.
Contours are plotted by an interval of 1.5\sgm\ starting from the 2\sgm\ level.  
The dashed-curves are the Keplerian rotation curves for the 
enclosed masses of $3\times 10^6$ \Msun\ 
assuming rotation velocity of 65 \kms\ at radii of 3\,pc (see text). 
The vertical red, green, and blue bars at the right-hand side indicate 
velocity ranges as shown in Figure \ref{fig:sp}. 
\label{fig:pv}}
\end{figure}

\begin{figure}[b!]
\includegraphics[angle=0,scale=.29]{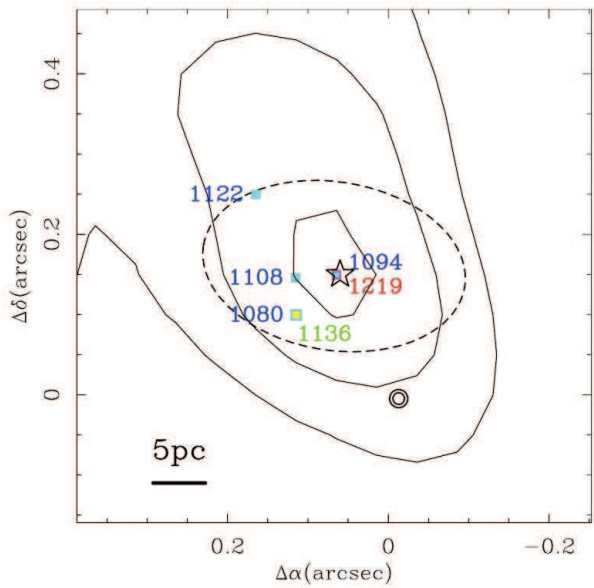}
\vspace{0truemm}
\caption{Comparison of the peak pixel position of 
the 689\,GHz continuum (star) 
and those obtained from the CO (6--5) velocity channel maps. 
The light-blue, yellow, and magenta squares with values of  
their LSR velocities
show the peak pixel positions of the blue, green, 
and red components shown 
in Figure \ref{fig:sp}, respectively. 
Note that all the squares shown here have S/N-ratio of higher than 4, 
as shown by the filled squares in Figure \ref{fig:chmaps}. 
The double-colors squares, 
the light-blue and yellow one with the double-labels of 
1080 and 1136 indicate the two velocity channels peak at the same position. 
Similarly the light-blue filled square enclosed by the magenta square labeled with
1094 and 1219 show the peak position of the two velocity components.
\label{fig:pixpos_summary}}
\end{figure}

\begin{figure}[ht!]
\includegraphics[angle=0,scale=.49]{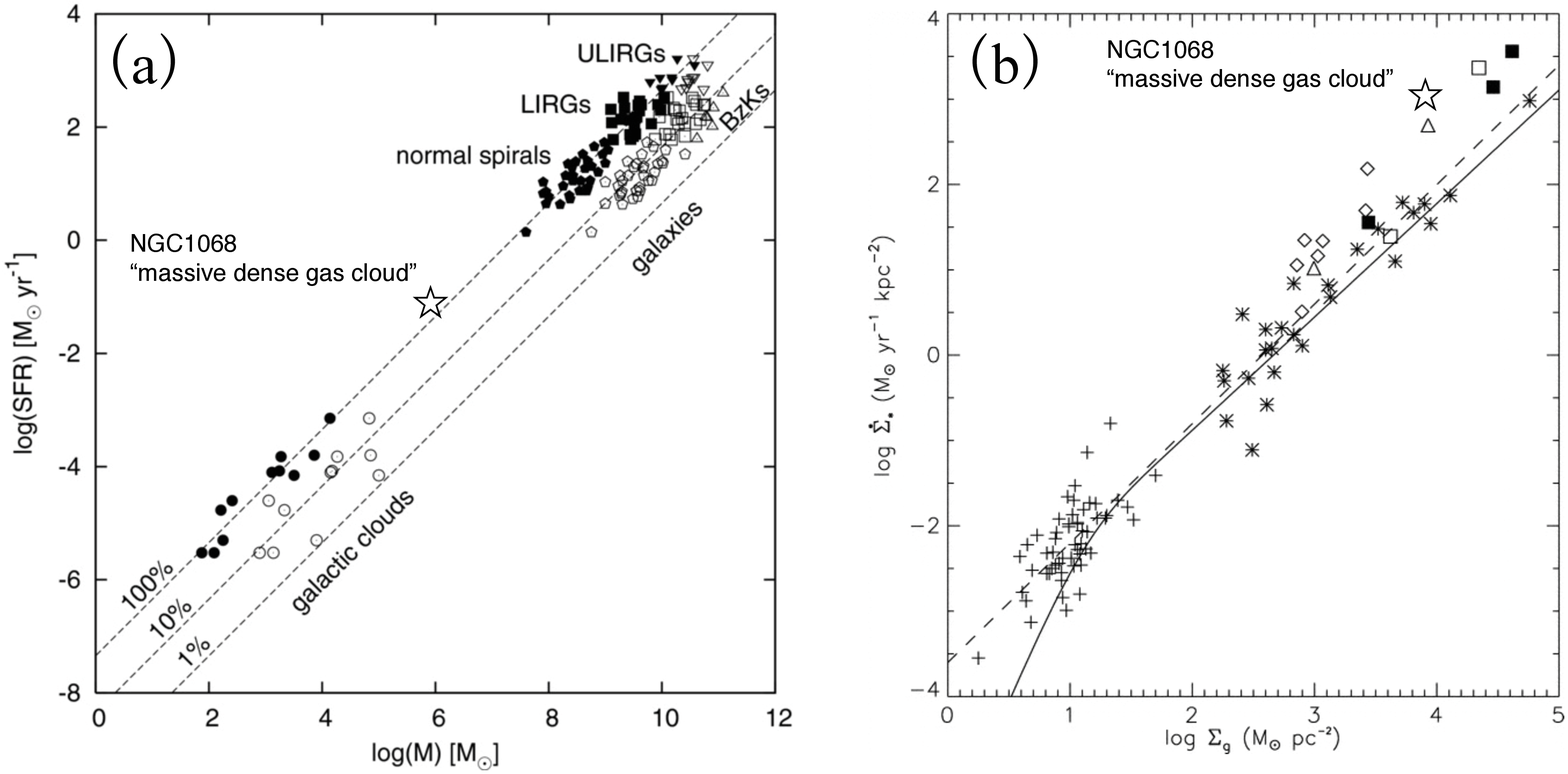}
\vspace{0truemm}
\caption{Comparisons of star formation properties of the
``massive dense gas cloud" (star) with the galactic and
extra-galactic clouds in the diagrams of 
(a) star formation rate (SFR) vs. gas mass, $M_\mathrm{gas}$, and 
(b) star formation rate density $\dot{\Sigma}_\mathrm{g}$ (\Msun\ yr$^{-1}$ kpc$^{-2}$) vs.
surface density of gas, $\Sigma_\mathrm{g}$ (\Msun\  pc$^{-2}$).
The plots of (a) and (b) are taken from 
\citet{Lada12} and \citet{KM05}, respectively, with permissions of the AAS.
The percentages labeled with the dashed lines in (a) indicate
dense gas fraction, $f_\mathrm{d}$ [see \citet{Lada12} for detail].
The dashed- and solid lines in (b) are the best-fit curve by \citet{K98}
and the model one by \citet{KM05} [see \citet{KM05} for detail].
\label{fig:SFR}}
\end{figure}

We sincerely acknowledge the anonymous referee whose
comments significantly helped to improve quality of our analysis and discussion.
The authors sincerely acknowledge Charles Lada, Mark Krumholz, and the Copyright \& Permissions Team of
the AAS journal for their kind permission to use their figures in this work (Figure \ref{fig:SFR}).
We would also like to thank Michael R. Blanton for providing us with his SDSS color composite image of NGC\,1068 shown in 
Figure\,\ref{fig:guidemap}a and Fumi Egusa for her generous support in handling the CO (6--5) image data. 
This work was financially supported in part by JSPS (YT; 23244041 and 16H02166).
This paper makes use of the ALMA data of ADS/JAO.ALMA$\#$2011.0.00083.S. 
ALMA is a partnership of ESO (representing its member states), NSF (USA) and NINS (Japan), 
together with NRC (Canada), NSC and ASIAA (Taiwan), and KASI (Republic of Korea), 
in cooperation with the Republic of Chile. 
The Joint ALMA Observatory is operated by ESO, AUI/NRAO and NAOJ.

\clearpage

\clearpage

\clearpage

\clearpage

\clearpage

\clearpage

\clearpage
\begin{center}
\begin{table}
\tbl{Properties of the ``massive dense gas cloud" in the nuclear region of NGC\,1068}{%
\begin{tabular}{@{}cc@{\qquad}ccc@{}}  
\hline\noalign{\vskip3pt} 
\multicolumn{1}{c}{Property} & Symbol & Value & \multicolumn{1}{c}{Unit} & Section \\  [2pt] 
\hline\noalign{\vskip3pt} 
R.A.  &  $\alpha_\mathrm{J2000}$  & 02:42:40.714 & h:m:s &  \ref{ss:ContPeak} \\
Decl. & $\delta_\mathrm{J2000}$ & $-00$:00:47.79 & \timeform{D}: \timeform{'}: \timeform{"} &  \ref{ss:ContPeak} \\
Size    &  2\Reff\       & $\sim\,5$ & pc &  \ref{ss:DP}, \ref{ss:SFandGproperty} \\
Temperature\footnotemark[$*$] & $T_\mathrm{d},\,T_\mathrm{gas}$ & 50 -- 70 & K & \ref{ss:SFandGproperty} \\
Gas plus dust mass & $M_\mathrm{gas}$ & $(5\pm 3)\times 10^5$ & \Msun\ & \ref{ss:SFandGproperty} \\
Column density & $\langle N_\mathrm{H_2}\rangle$ & $(7\pm 2)\times 10^{23}$ & cm$^{-2}$ & \ref{ss:SFandGproperty} \\
Surface gas density & $\log\,\Sigma_\mathrm{g}$ & $ 3.9\pm 0.2$ & \Msun\ pc$^{-2}$ & \ref{ss:nature} \\
Volume density & $n_\mathrm{H_2}$ & $\sim 1\times 10^5$ & cm$^{-3}$ & \ref{ss:SFandGproperty} \\
Bolometric luminosity & $L_\mathrm{bol}$ & (0.4 -- 4)$\times 10^8$ & \Lsun\ & \ref{ss:SFandGproperty} \\
Stellar mass & $M_\ast$ & a few $\times 10^4$ & \Msun\ & \ref{ss:SFandGproperty} \\
Star formation rate & SFR & (0.4 -- 3.2)$\times 10^{-1}$ & \Msun\ yr$^{-1}$ & \ref{ss:SFandGproperty} \\
Star formation rate density & $\log\,\dot{\Sigma}_\mathrm{g}$ & $2.9\pm 0.2$ & \Msun\ yr$^{-1}$ kpc$^{-2}$& \ref{ss:nature} \\
(Mach number)$^2$ & $M_\mathrm{vir}/M_\mathrm{vir}^\mathrm{thm}$ & (2 -- 3)$\times 10^3$ & $\cdot\cdot\cdot$ & \ref{ss:SFandGproperty}, \ref{ss:origin}\\
\hline\noalign{\vskip3pt} 
\end{tabular}}
\label{tbl:Ps}
\begin{tabnote}
\hangindent6pt\noindent
\hbox to6pt{\footnotemark[$*$]\hss}\unskip%
 Assumption.
\end{tabnote}
\end{table}
\end{center}
%

\clearpage


\begin{thebibliography}{}
\bibitem[Andr\'e et al.(1996)]{Andre96} Andr\'e, P., Ward-Thompson, D., \& Motte, F.\ 1996, \aap, 314, 625 

\bibitem[Antonucci \& Miller(1985)]{AM85} Antonucci, R. R. J., \& Miller, J. S.\ 1985, \apj, 297, 621 
\bibitem[Beckwith et al.(2000)]{Beckwith00} Beckwith, S.~V.~W., Henning, T., \& Nakagawa, Y.\ 2000, Protostars and Planets IV, 533 
\bibitem[Beltr{\'a}n et al.(2005)]{Beltran05} Beltr{\'a}n, M.~T., Cesaroni, R., Neri, R., et al.\ 2005, \aap, 435, 901 

\bibitem[Bodenheimer(1995)]{Bodenheimer95} Bodenheimer, P.\ 1995, \araa, 33, 199 
\bibitem[Cecil et al.(1990)]{Cecil90} Cecil, G., Bland, J., \& Tully, R.~B.\ 1990, \apj, 355, 70 
\bibitem[Cecil et al.(2002)]{Cecil02} Cecil, G., et al.\ 12002, \apj, 568,  627
\bibitem[Crutcher(2012)]{Crutcher12} Crutcher, R.~M.\ 2012, \araa, 50, 29

\bibitem[de Vaucouleurs et al.(1991)]{dV91} de Vaucouleurs, G., de Vaucouleurs, A., Corwin, H.~G., Jr., et al.\ 1991, Third Reference Catalogue of Bright Galaxies.~Volume I: Explanations and references.~ Volume II: Data for galaxies between 0$^{h}$ and 12$^{h}$.~ Volume III: Data for galaxies between 12$^{h}$ and 24$^{h}$., by de Vaucouleurs, G.; de Vaucouleurs, A.; Corwin, H.~G., Jr.; Buta, R.~J.; Paturel, G.; Fouqu{\'e}, P..~Springer, New York, NY (USA), 1991, 2091 p., ISBN 0-387-97552-7, Price US 198.00. ISBN 3-540-97552-7, Price DM 448.00. ISBN 0-387-97549-7 (Vol. I), ISBN 0-387-97550-0 (Vol. II), ISBN 0-387-97551-9 (Vol. III)., I,  


\bibitem[Dickman(1978)]{Dickman78} Dickman, R.~L.\ 1978, \apjs, 37, 407 
\bibitem[Furuya et al.(2009)]{RSF09} Furuya, R.~S., Kitamura, Y., \& Shinnaga, H.\ 2009, \apjl, 692, L96 
\bibitem[Furuya et al.(2011)]{RSF11} Furuya, R.~S., Cesaroni, R., \& Shinnaga, H.\ 2011, \aap, 525, A72 
\bibitem[Furuya et al.(2014)]{RSF14} Furuya, R.~S., Kitamura, Y., \& Shinnaga, H.\ 2014, \apj, 793, 94 
\bibitem[Gallimore et al.(1996a)]{G96_5GHzcont} Gallimore, J.~F., Baum, S.~A., O'Dea, C.~P., \& Pedlar, A.\ 1996a, \apj, 458, 136 
\bibitem[Gallimore et al.(1996b)]{G96_H2O} Gallimore, J.~F., Baum, S.~A., O'Dea, C.~P., Brinks, E., \& Pedlar, A.\ 1996b, \apj, 462, 740 
\bibitem[Gallimore et al.(1996c)]{G96_MERLINt} Gallimore, J.~F., Baum, S.~A., O'Dea, C.~P., \& Pedlar, A.\ 1996c, \apj, 458, 136 
\bibitem[Gallimore et al.(2001)]{G01} Gallimore, J.~F., Henkel, C., Baum, S.~A., et al.\ 2001, \apj, 556, 694 
\bibitem[Gallimore et al.(2004)]{G04} Gallimore, J.~F., Baum, S.~A., \& O'Dea, C.~P.\ 2004, \apj, 613, 794 

\bibitem[Garc{\'{\i}}a-Burillo et al.(2014a)]{GB14a} Garc{\'{\i}}a-Burillo, S., Combes, F., Usero, A., et al.\ 2014, \aap, 567, A125 

\bibitem[Garc{\'{\i}}a-Burillo et al.(2014b)]{GB14b} Garc{\'{\i}}a-Burillo, S., Fuente, A., Hunt, L. K., et al.\ 2014, \aap, 570, A28 

\bibitem[Garc{\'{\i}}a-Burillo et al.(2016)]{GB16} Garc{\'{\i}}a-Burillo, S., Combes, F., Ramos, A., et al.\ 2016, \apj, 823, L12 


\bibitem[Greenhill et al.(1996)]{Greenhill96} Greenhill, L.~J., Gwinn, C.~R., Antonucci, R., \& Barvainis, R.\ 1996, \apjl, 472, L21 

\bibitem[Habe \& Ohta(1992)]{Habe92} Habe, A., \& Ohta, K.\ 1992, \pasj, 44, 203 

\bibitem[Hasegawa et al.(1994)]{Hasegawa94} Hasegawa, T., Sato, F., Whiteoak, J.~B., \& Miyawaki, R.\ 1994, \apjl, 429, L77 

\bibitem[Heckman \& Best(2014)]{HB14} Heckman, T.,  \& Best, P.\ 2014, \araa, 52, 589 
\bibitem[Hopkins et al.(2008)]{Hopkins08} Hopkins, P.~F., Hernquist, L., Cox, T.~J., \& Kere{\v s}, D.\ 2008, \apjs, 175, 356 

\bibitem[Imanishi et al.(2016)]{Ima16} Imanishi, M., Nakanishi, K.,  \& Izumi, T.\ 2016, \apj, 822, L10 

\bibitem[Inoue \& Fukui(2013)]{IF13} Inoue, T., \& Fukui, Y.\ 2013, \apjl, 774, L31 

\bibitem[Inutsuka et al.(2015)]{SI15} Inutsuka, S.-i., Inoue, T., Iwasaki, K., \& Hosokawa, T.\ 2015, \aap, 580, A49 

\bibitem[Izumi et al.(2016)]{Izu16} Izumi, T.,  Nakanishi, K., Imanishi, M., \& Kohno, K. 2016, \mnras, 459, 3629 

\bibitem[Kennicutt(1998)]{K98} Kennicutt, R.~C., Jr.\ 1998, \araa, 36, 189 


\bibitem[Kennicutt \& Evans(2012)]{KE12} Kennicutt, R.~C., \& Evans, N.~J.\ 2012, \araa, 50, 531 

\bibitem[Khan et al.(2012)]{Khan12} Khan, F.~M., Preto, M., Berczik, P., et al.\ 2012, \apj, 749, 147 

\bibitem[Khachikian \& Weedman(1974)]{KW74} Khachikian, E. Y., \& Weedman, D. W. D.1974, \apj, 192, 581 

\bibitem[Klaas et al.(2001)]{Klaas01} Klaas, U., Haas, M., M{\"u}ller, S.~A.~H., et al.\ 2001, \aap, 379, 823 

\bibitem[Kormendy \& Ho(2013)]{KH13} Kormendy, J., \& Ho, L.~C.\ 2013, \araa, 51, 511 

\bibitem[Krumholz \& McKee(2005)]{KM05} Krumholz, M.~R., \& McKee, C.~F.\ 2005, \apj, 630, 250 

\bibitem[Kurtz et al.(2000)]{Kurtz00} Kurtz, S., Cesaroni, R., Churchwell, E., Hofner, P., \& Walmsley, C.~M.\ 2000, Protostars and Planets IV, 299 


\bibitem[Kulier et al.(2015)]{Kulier15} Kulier, A., Ostriker, J.~P., Natarajan, P., Lackner, C.~N., \& Cen, R.\ 2015, \apj, 799, 178 
\bibitem[Lada et al.(2012)]{Lada12} Lada, C.~J., Forbrich, J., Lombardi, M., \& Alves, J.~F.\ 2012, \apj, 745, 190 

\bibitem[Liu et al.(2016)]{Liu16} Liu, J., Eracleous, M., \& Halpern, J.~P.\ 2016, \apj, 817, 42 

\bibitem[Lopez-Rodriguez et al.(2016)]{LR16} Lopez-Rodriguez, E., Packham, C., Roche, P.~F., et al.\ 2016, \mnras, 458, 3851 

\bibitem[Mezcua et al.(2015)]{Mezcua15} Mezcua, M., Prieto, M.~A., Fern{\'a}ndez-Ontiveros, J.~A., et al.\ 2015, \mnras, 452, 4128 

\bibitem[Miyoshi et al.(1995)]{Miyoshi95} Miyoshi, M., Moran, J., Herrnstein, J., et al.\ 1995, \nat, 373, 127 

\bibitem[Mihos \& Hernquist(1994)]{MH94} Mihos, C. J., \& Hernquist, L.\ 1994, \apj, 425, L13 
\bibitem[Mulia et al.(2016)]{Mulia16} Mulia, A. J.,Rupali, C.,  \& Whitmorei, B. C.\ 2016, arXiv:1607.03577 

\bibitem[Murayama \& Taniguchi(1997)]{MT97} Murayama, T., \& Taniguchi, Y.\ 1997, \pasj, 49, L13 

\bibitem[Papaloizou \& Pringle(1984)]{PP84}Papaloizou, J. C. B., , \& Pringle, J. E.\ 1984, \mnras, 208, 721


\bibitem[Planck Collaboration et al.(2011b)]{PC11b} Planck Collaboration, Abergel, A., Ade, P.~A.~R., et al.\ 2011, \aap, 536, A25 

\bibitem[Preibisch et al.(1993)]{Preibisch93} Preibisch, T., Ossenkopf, V., Yorke, H.~W., \& Henning, T.\ 1993, \aap, 279, 577 
\bibitem[Rees(1984)]{Rees84} Rees, M.~J.\ 1984, \araa, 22, 471 
\bibitem[Sanders \& Mirabel(1996)]{Sanders96} Sanders, D.~B., \& Mirabel, I.~F.\ 1996, \araa, 34, 749 
\bibitem[Seyfert(1943)]{Sey43} Seyfert, C. K. 1943, \apj, 97, 28

\bibitem[Shaya(1994)]{Shaya94} Shaya, E. J., et al. \ 1994, \apj, 107, 1675 
\bibitem[Shioya(2001)]{Shioya01} Shioya, Y., Taniguchi, Y., \& Trentham, N.  2001, \mnras, 321, 11


\bibitem[Schinnerer et al.(2000)]{Sch00} Schinnerer, E., Eckart, A., Tacconi, L.~J., Genzel, R., \& Downes, D.\ 2000, \apj, 533, 850
\bibitem[Simkin et al.(1980)]{Sim80} Simkin, S. M., Su, H. J., \& Schwarz.M. P.\ 1980, \apj,237. 404 

\bibitem[Storchi-Bergamnn et al.(1996)]{SB96} Storchi-Bergamnn, T. et al. 1996, \apj, 472, 83 
\bibitem[Storchi-Bergamnn et al.(2012)]{SB12} Storchi-Bergamnn, T., et al.  2012, \apj, 755, 87 


\bibitem[Taniguchi \& Wada(1996)]{TW96} Taniguchi, Y., \& Wada, K.\ 1996, \apj, 469, 581 
\bibitem[Taniguchi(1999)]{YT99} Taniguchi, Y. 1999, \apj, 524, 65 
\bibitem[Taniguchi(2013)]{YT13} Taniguchi, Y. 2013, Galaxy Mergers in an Evolving Universe, 477, 265 
\bibitem[van der Tak et al.(2007)]{vdTak07} van der Tak, F.~F.~S., Black, J.~H., Sch{\"o}ier, F.~L., Jansen, D.~J., \& van Dishoeck, E.~F.\ 2007, \aap, 468, 627 

\bibitem[Vale et al.(2012)]{Vale12} Vale, T.~B., Storchi-Bergmann, T., \& Barbosa, F.~K.~B.\ 2012, AGN Winds in Charleston, 460, 164 

\bibitem[Wang et al.(2012)]{Wang12} Wang, J., Fabbiano, G., Karovska, M., Elvis, M., \& Risaliti, G.\ 2012, \apj, 756, 180 

\bibitem[Whitmore et al.(1993)]{Whitmore93} Whitmore, B. C., et al. \ 1993, \aj, 106, 1354 
\bibitem[Wilson et al.(1991)]{Wilson91} Wilson, A. S., ,Helfer, T. T.,  Haniff, C. A. \& Ward, M. J. 1993, \apj, 381, 79



\bibitem[Zhou et al.(1993)]{Zhou93} Zhou, S., Evans, N.~J., II, Koempe, C., \& Walmsley, C.~M.\ 1993, \apj, 404, 232 
\end{thebibliography}
\end{document}